\documentclass[twocolumn,english,aps,pre,showpacs]{revtex4}
\usepackage{srcltx}
\usepackage{amsmath}
\usepackage{graphicx}

\begin{document}

\title{The propagation of Elastic Waves in Granular
Solid Hydrodynamics}

\author{Michael Mayer}

\author{Mario Liu}

\affiliation{Institut f\"{u}r Theoretische Physik,
Universit\"{a}t T\"{u}bingen, 72076 T\"{u}bingen, Germany, EU}

%\email{m.mayer@ or mliu@uni-tuebingen.de }

\date{\today}
\begin{abstract}
The anisotropic, stress-dependent velocity of elastic
waves in glass beads -- as observed by Y.~Khidas and
X. Jia, see~[Phys. Rev. E, 81:021303, Feb. 2010] -- is
shown to be well accounted for by ``granular solid
hydrodynamics,'' a broad-range macroscopic theory of
granular behavior. As the theory makes no reference to
fabric anisotropy, the influence of which on sound is
in doubt.
\end{abstract}


\maketitle

\section{Introduction }

The anisotropic spectrum of elastic waves in granular
media is studied in~\cite{Jia2009}. Wave velocity was
measured under varying stress, in glass bead samples
that have undergone two different preparations: {\em
Rain deposition} that consists of pouring the glass
beads into the acoustic cell through a mesh; and {\em
de-compaction}, implying the gentle motion of a
horizontal grid through the bead pack, from the bottom
to the top, after the deposition. As the velocity was
seen to depend on the preparation, and because it is
known from numerical simulations that the rain
deposition creates an anisotropic distribution of the
contact angles, with two preferred directions oriented
roughly at $30^\circ$ from the vertical, this was
taken as evidence of the influence of fabric
anisotropy.

However, also reported was the fact that the
preparations lead to different densities. So a crucial
question is whether the density alone suffices to
explain the observed difference in the velocity. Only
if this is not the case, would the experimental
evidence imply that fabric anisotropy exerts
independent influence, that it is in fact a state
variable -- in addition to, and independent from,
others such as the stress, density, velocity, and the
granular temperature. Generally speaking, of course,
fabric anisotropy could well behave like many other
local or microscopic variables, is different in
different circumstances but, being a dependent
quantity, would be given if the complete set of state
variables is.

\begin{figure}[b]
\includegraphics[scale=.85]{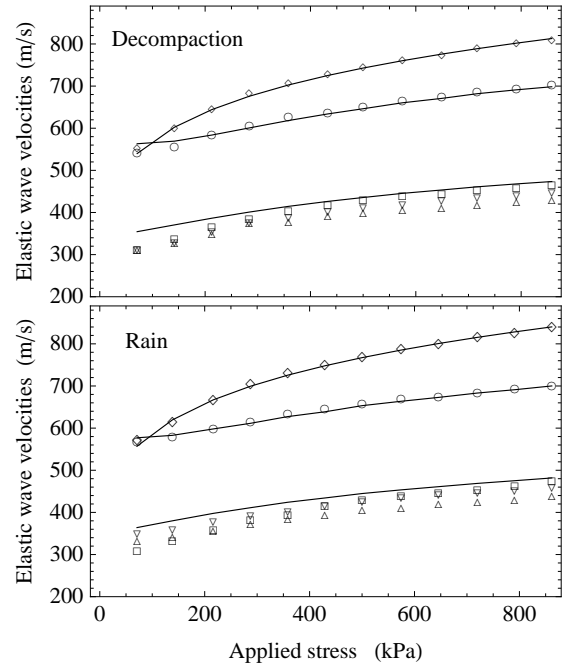}
\caption{\label{fig1}Symbols represent measured wave
velocities versus the applied stress $\sigma_{33}$, as
reported by Khidas and Jia, for two different
preparations, decompaction and rain, corresponding
respectively to the packing ratio of $\phi=0.606$ and
$\phi=0.643$. (The geometry is triaxial, the preferred
direction along the vertical is denoted as 3.) More
specifically, rhombuses are vertical compressional
waves $c_v$; circles horizontal ones $c_h$. The
remaining three are shear waves, with varying
direction and polarization, see~\cite{Jia2009}. The
lines are theoretical results, obtained by employing
{\sc gsh} with density-dependent elastic coefficients,
but excluding any possible independent influences from
preparations. In the employed approximation of two
elastic coefficients, assuming uniaxiality, {\sc gsh}
yields the same velocity for all three shear waves --
a justified approximation, as their difference is
clearly small.}
\end{figure}

Fig.~\ref{fig1} demonstrates that the velocity
difference can indeed be explained by the density
difference. Although there is some residual
discrepancy between theory and observation, many
possible reasons remain, making the case for fabric
anisotropy less than cogent: The calculation was
carried out under simplifying assumptions, and the
experiment contains deviation from ideal conditions.
The employed theory, {\sc gsh} (for granular solid
hydrodynamics)~\cite{Liu2009}, contains only two
elastic coefficients, leaving the energy independent
of the third strain invariant, see
section~\ref{sec:Anisotropy-of-granular}. On the
experimental side, possibly neither the density nor
the static stress is uniform, as assumed. The latter
is mainly a result of the elliptical geometry, and the
inhomogeneity in density should especially be
dependent on preparation.

Having adjusted the parameters such that the density
difference from the two preparations suffices to explain
the velocity difference in the compressional waves,
Khidas and Jia find that the discrepancy in the shear
wave velocities remains, and depends on the preparation,
see their Fig. 8. Accepting this as an evidence for the
independent influence of fabric anisotropy, however,
presumes complete confidence in the density dependence of
{\sc emt}, the theory employed in~\cite{Jia2009} to
account for the wave velocities. {\sc gsh} sports a more
realistic density
dependence~\cite{Liu2009,jiang,brauer,krimer}, and the
same discrepancy is small. In addition, the neglected
elastic coefficients associated with the third strain
invariant should also depend on the density, further
reducing the discrepancy.

We derive the elastic modes from {\sc gsh} in
section~\ref{sec:Elastic-sound-modes}, and apply the
results to the experimental circumstances
of~\cite{Jia2009} in
section~\ref{sec:Anisotropy-of-granular}.

\section{Elastic modes in the { GSH} theory
\label{sec:Elastic-sound-modes}}

GSH theory is a hydrodynamic theory meant to account for
a wide spectrum of granular behavior,
see~\cite{Liu2009,jiang,brauer,krimer} for derivation and
explanation. Its variables consist of the density $\rho$,
momentum $\rho v_i$, the elastic strain $u_{ij}$, and two
entropies, the true and the granular one, $s, s_g$. To
consider elastic waves, we only need the equations
\begin{eqnarray*}
\partial_{t}\left(\rho v_{i}\right)+\nabla_{j}\left(\sigma_{ij}
+\rho v_{i}v_{j}\right) & = & 0,\\
{\partial_{t}u_{ij}}-\left(1-\alpha\right)v_{ij}
+\frac{u_{ij}^{0}}{\tau}+\frac{u_{ll}\delta_{ij}}{\tau_{1}}
& = & 0,
\end{eqnarray*}
with the stress given as
$\sigma_{ij}=\left(1-\alpha\right)\pi_{ij}
-\zeta_{g}v_{ll}\delta_{ij}-\eta_{g}v_{ij}^{0}$,
$\pi_{ij}\equiv-\partial w/\partial u_{ij}$, and $w$
denoting the elastic energy. The coefficients $\alpha$
and $\tau^{-1},\tau_1^{-1}$ account respectively for
the typical granular phenomena of stress softening and
strain relaxation that give rise to plasticity. These
coefficients and the viscosities $\eta_g, \zeta_g$ are
all functions of the granular temperature and vanish
for $T_g\to0$. For small wave amplitudes, the granular
temperature $T_{g}$ is also small, making all these
effects negligible, and granular media essentially
elastic. This is the limit we shall consider here,
while postponing the account of wave dissipation from
viscosity and plasticity to a forthcoming paper.
Taking $\alpha, \tau^{-1}, \tau_1^{-1}, \eta_g,
\zeta_g=0$, and
$u_{ij}=\frac{1}{2}\left(\partial_{i}U_{j}
+\partial_{j}U_{i}\right)$, we have
\begin{equation}
\rho\ddot{U}_{i}+M_{ijkl}\nabla_{jk}U_{l}=0,
\label{eq:waveequation}\end{equation} a wave equation for
the displacements $U_{i}$, with the stress-dependent
stiffness tensor~\cite{jiang},
$M_{ijkl}=\partial^2w/\partial u_{ij}\partial u_{kl}$. We
take the granular elastic energy to be
\begin{equation}
w=\mathcal{B}\sqrt{\Delta}\left({2}
\Delta^{2}/{5}+{u_{s}^{2}}/{\xi}\right),\label{eq:energy}
\end{equation}
with $\Delta\equiv-u_{ll}$, $u_{s}^{2}\equiv
u_{ij}^{0}u_{ij}^{0}$, and the superscript $^{0}$
denoting the traceless part of any tensor, eg.
$u_{ij}^{0}\equiv u_{ij}+\delta_{ij}\Delta/3$. The two
elastic coefficients are $\mathcal{B}$ and $\xi$,
where  $\xi$ is a constant, and $\cal B$ density
dependent,
\begin{equation}
\mathcal{B}=\mathcal{B}_{0}
\left(\frac{\phi-\phi_{lp}^{*}}{\phi_{cp}
-\phi}\right)^{\frac{3}{20}}, \label{eq:B}\end{equation}
where $\phi\equiv\rho/\rho_{G}$ is the packing ratio,
with $\phi_{cp},\phi_{lp}$ respectively the ratio of
closest and loosest packing, and $\phi_{lp}^{*}$ given
by $\phi_{lp}=(11\phi_{cp}+9\phi_{lp}^{*})/20$.

Inverting the stress-strain relation
$\pi_{ij}=\pi_{ij}(u_{kl})$, taking $P\equiv
\pi_{ll}/3$ and $\pi_{s}^{2}\equiv
\pi_{ij}^{0}\pi_{ij}^{0}$, we express the elastic
strain as
\begin{eqnarray}%\label{}
2L=1+\sqrt{1-\xi\pi_{s}^{2}/(2P^{2})}
\\\Delta=(LP/\mathcal{B})^{\frac{2}{3}},
\quad u_{s}=\pi_{s}/(2\mu),
\\\label{111}
u_{ij}^{0}=-\frac{\pi_{ij}^{0}}{2\mu},\quad
\mu=\frac{\cal B}{\xi}
\left(\frac{LP}{\mathcal{B}}\right)^{{1}/{3}}.
\end{eqnarray}

For uniform stress and strain, in the principal axes of
both the stress and strain [see the first of
Eqs~(\ref{111})], $\pi_{ij}=\pi_{(i)}\delta_{ij}$, the
stiffness tensor has the simple form
\begin{eqnarray}\nonumber
M_{ijkl}&=&\frac{\mathcal{B}}{\xi}
\left(\frac{LP}{\mathcal{B}}\right)^{{1}/{3}}
\left(a_{(ik)}\delta_{ij}\delta_{kl}-
\delta_{ik}\delta_{jl}-\delta_{il}\delta_{jk}\right),
\\\label{eq:stiffness}
a_{ik}&\equiv&\left(\frac{\xi\pi_{s}}{4LP}\right)^{2}+\frac{4-9\xi}{6}
-\frac{\xi(\pi_{i}^{0}+\pi_{k}^{0})}{2LP},
\end{eqnarray}
where there is no summation over bracketed subscripts.
So,  for instant
\[M_{1122}=\frac{\cal B}\xi\left(\frac{LP}{\mathcal{B}}
\right)^{\frac{1}{3}}
\left[\left(\frac{\xi\pi_{s}}{4LP}\right)^{2}+\frac{4-9\xi}{6}
+\frac{\xi(\pi_{1}^{0}+\pi_{2}^{0})}{2LP}\right].\]
A plane wave {\em ansatz} $\sim exp(\omega
t+\vec{k}\vec{x})$ to eq.(\ref{eq:waveequation})
results in the eigenvalue problem
$\left(K_{il}-\lambda\delta_{il}\right)U_{l}=0$, where
\begin{eqnarray*}%\label{}
K_{il}\equiv(\beta_{(i)}+\beta_{(l)})n_{i}n_{l}/2-\delta_{il},
\\
c^{2}\equiv\omega^{2}/k^{2}=-\lambda\mathcal{B}
(LP/\mathcal{B})^{\frac{1}{3}}/(\rho\xi),
\\
\beta_{i}\equiv1-Q-\frac{\xi\pi_{i}^{0}}{LP}, \quad
Q\equiv\frac{3\xi}2+\frac43-\left(\frac{\xi\pi_{s}}{4LP}\right)^{2},
\end{eqnarray*}
%,
and $n_{i}=k_{i}/k$. Eigenvalues $\lambda$ yield the
velocity of propagation $c$, and the corresponding
eigenvectors $U$ the polarization. The general
solution is given by the eigenvalues $\lambda_{1}=-1$
and
$\lambda_{2,3}=\frac{1}{2}\left(\eta\mp\sqrt{\zeta}-2\right)$,
where $\eta\equiv\sum_{i=1}^{3}\beta_{i}n_{i}^{2}$ and
$\zeta\equiv\sum_{i=1}^{3}\beta_{i}^{2}n_{i}^{2}$.

As long as $\beta_{1}\neq\beta_{2}\neq\beta_{3}$, and
the wave vector is not along one of the principal
directions, the associated displacements are
\begin{eqnarray}%multline
U_{1}&=&\begin{pmatrix}n_{2}n_{3}(\beta_{2}-\beta_{3})\\
n_{3}n_{1}(\beta_{3}-\beta_{1})\\
n_{1}n_{2}(\beta_{1}-\beta_{2})\end{pmatrix},\\\nonumber
U_{2,3}&=&\begin{pmatrix}n_{1}(\zeta\mp\sqrt{\zeta}(\beta_{1}+\beta_{1})+\beta_{1}\beta_{1})\\
n_{2}(\zeta\mp\sqrt{\zeta}(\beta_{1}+\beta_{2})+\beta_{1}\beta_{2})\\
n_{3}(\zeta\mp\sqrt{\zeta}(\beta_{1}+\beta_{3})+\beta_{1}\beta_{3})
\end{pmatrix}.\label{eq:64}\end{eqnarray}
In the case of propagation along principal directions
the tensor $K$ is diagonal from the very beginning,
and we have, eg., for $n_{1}=1$,
$\lambda_{1}=\beta_{1}-1$ and $\lambda_{2,3}=-1$ with
$U_{1}=(1,0,0)$, $U_{2}=(0,1,0)$ and $U_{3}=(0,0,1)$.
Thus, plane waves traveling along the principal
directions of the static stress field are either pure
longitudinal or transversal modes.

For an isotropic background stress we have
$\pi_{i}^{0}=0$, and $\beta_{1}=\beta_{2}=\beta_{3}$, and
the compressional wave velocity shows the typical
$P^{\frac{1}{6}}$ Hertz-scaling,
\begin{equation}
c_{p}=\sqrt{\left(\frac{3}{2}\xi+
\frac{4}{3}\right)\frac{1}{\xi\phi}
\frac{\mathcal{B}}{\rho_{G}}}
\sqrt[6]{\frac{P}{\mathcal{B}}}.
\label{eq:cpiso}\end{equation}
For a uniaxial state, with 3 as the preferred
direction,  we have $\beta_{1}=\beta_{2}$ and
$\lambda_{1}=-1$,
$\lambda_{2,3}=\frac{1}{2}\left(\eta\mp\sqrt{\zeta}-2\right)$,
with
$\eta\equiv\left(\beta_{1}-\beta_{3}\right)n_{1}^{2}+\beta_{3}$,
$\zeta\equiv\left(\beta_{1}^{2}-\beta_{3}^{2}\right)
n_{1}^{2}+\beta_{3}^{2}$. Depending on $n_1$ alone,
$\lambda$ (and therefore also the velocities $c$)
possesses rotational symmetry. The compressional wave
velocity along 3, or $n_{1}=0$, is given by
\begin{equation}
c_{v}=\sqrt{\left(Q+\frac{\pi_{3}^{0}}{LP}\right)
\frac{1}{\xi\phi}\frac{\mathcal{B}}
{\rho_{G}}}\sqrt[6]{\frac{LP}{\mathcal{B}}}.
\label{eq:cpuni}\end{equation}
The lateral compressional velocity, $n_{1}=1$, is
\begin{equation}
c_{h}=\sqrt{\left(Q+\frac{\pi_{1}^{0}}{LP}\right)
\frac{1}{\xi\phi}\frac{\mathcal{B}}{\rho_{G}}}
\sqrt[6]{\frac{LP}{\mathcal{B}}}.
\label{eq:cpuni1}\end{equation}
Only for vanishing shear, $\pi_{i}^{0},\pi_{s}\to0$,
$L\to1$, do both formulas reduce to
Eq~(\ref{eq:cpiso}). Otherwise, the $\sqrt[6]P$-
Hertz-scaling is not valid. The two shear wave modes
for each of the two directions all have the
eigenvalue, $\lambda=-1$, so
\begin{eqnarray} c_{s} & = &
\sqrt{\frac{1}{\xi\phi}\frac{\mathcal{B}}{\rho_{G}}}
\sqrt[6]{\frac{LP}{\mathcal{B}}}.\label{eq:cs}\end{eqnarray}

\section{Anisotropy of granular sound
\label{sec:Anisotropy-of-granular}}
These results are now applied to analyze the data
of~\cite{Jia2009}. From
Eqs~(\ref{eq:cpuni},\ref{eq:cpuni1}), we first
calculate, for the uniaxial case, the ratio of
vertical and horizontal compressional velocities,
\begin{equation}
\left(\frac{c_{v}}{c_{h}}\right)^{2}=
\frac{LPQ+\pi_{3}^{0}}{LPQ+\pi_{1}^{0}},
\end{equation}
which clearly does not depend on $\mathcal{B}$, only on
$\xi$, determined as $\xi={3}/{4}$ from the measured
velocities of both preparations, rain and decompaction.
Clearly, $\xi$ does not depend much on the density, as
assumed. Then, taking $\phi_{lp}=0.555$ and
$\phi_{cp}=0.664$, any data point will fix
$\mathcal{B}_{0}$ as $5.1$ GPa.  The result is
Fig~\ref{fig1}. In fact, data points from compressional
waves were used to fix $\xi$ and $\mathcal{B}_{0}$, so
the curves of shear wave velocities are free of fit
parameters.

The degenerate velocity for shear waves is a consequence
of the simple form given for the energy in
Eq~(\ref{eq:energy}). Generally speaking, there are three
strain invariants: $\Delta\equiv-\text{Tr}\,\hat u$,
$u_{s}^{2}\equiv\text{Tr}\, \hat
u^{2}-\frac{1}{3}\left(\text{Tr}\, \hat u\right)^{2}$,
and  $u_{III}\equiv\left(\text{Tr}\hat u\right)^{3}-$
$3\text{Tr}\,\hat u\,\text{Tr}\,\hat u^{2}+2\text{Tr}\,
\hat u^{3}$, of which only the first two enter the
expression of Eq~(\ref{eq:energy}). Including the third
invariant, we did find a discrimination of shear wave
velocities under static shear. [Note that although
$u_{III}$ is of third order, its energy contribution, say
$\sim u_{III}/\sqrt{\Delta}$, may well be of the same
order as the terms already in Eq~(\ref{eq:energy}).] We
are at present looking for a suitable extension of
Eq~(\ref{eq:energy}), taking into account in addition its
effect on the Coulomb yield.

\section{Summary}

We have shown that {\sc gsh}, employing only two
independent elastic coefficients, ${\cal B}_0=5.1$ GPa,
$\xi=3/4$, is capable of reproducing the experimental
date of~\cite{Jia2009} fairly well. Various possible
reasons for the remaining disagreement are given and
weighted, all rather more tangible than fabric
anisotropy. Moreover, the density dependence of ${\cal
B}$ and the density independence of $\xi$ is further
validated.

\end{document}